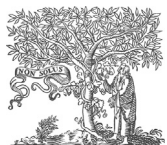
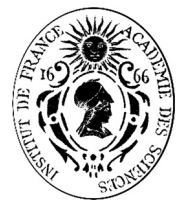

## Comptes Rendus Physique

www.sciencedirect.com

Condensed matter physics in the 21st century: The legacy of Jacques Friedel

# Jacques Friedel and the physics of metals and alloys

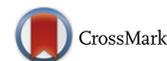

## *Jacques Friedel et la théorie des métaux et alliages*

Jacques Villain [a,*], Mireille Lavagna [b,c], Patrick Bruno [d]

[a] *Theory group, Institut Laue–Langevin, 38054 Grenoble cedex 9, France*
[b] *Université Grenoble Alpes, INAC–SPSMS, 38000 Grenoble, France*
[c] *CEA, INAC–SPSMS, 38000 Grenoble, France*
[d] *Theory group, European Synchrotron Radiation Facility, 38054 Grenoble cedex 9, France*

**ABSTRACT**

This is an introduction to the theoretical physics of metals for students and physicists from other specialities. Certain simple consequences of the Fermi statistics in pure metals are first addressed, namely the Peierls distortion, Kohn anomalies and the Labbé–Friedel distortion. Then the physics of dilute alloys is discussed. The analogy with nuclear collisions was a fruitful starting point, which suggested one should analyze the effects of impurities in terms of a scattering problem with the introduction of phase shifts. Starting from these concepts, Friedel derived a theory of the resistivity of alloys, and a celebrated sum rule relating the phase shifts at the Fermi level to the number of electrons in the impurity, which turned out to play a prominent role later in the context of correlated impurities, as for instance in the Kondo effect. Friedel oscillations are also an important result, related to incommensurate magnetic structures. It is shown how they can be derived in various ways: from collision theory, perturbation theory, self-consistent approximations and Green's function methods. While collision theory does not permit to take the crystal structure into account, which is responsible for electronic bands, those effects can be included in other descriptions, using for instance the tight binding approximation.

© 2015 The Authors. Published by Elsevier Masson SAS on behalf of Académie des sciences. This is an open access article under the CC BY-NC-ND license (http://creativecommons.org/licenses/by-nc-nd/4.0/).

**RÉSUMÉ**

Cet article est une introduction à la théorie électronique des métaux. Il s'adresse aux étudiants et aux physiciens non spécialistes. On commence par décrire certaines conséquences simples de la statistique de Fermi–Dirac dans les métaux purs, comme la distorsion de Peierls, les anomalies de Kohn et la distorsion de Labbé–Friedel. On discute ensuite la physique des alliages dilués. L'analogie avec le problème des collisions nucléaires fut un point de départ fructueux, qui amena à considérer l'effet des impuretés comme un problème de diffusion, dans lequel apparaissent les déphasages de l'onde électronique diffusée. Friedel élabora ainsi une théorie de la résistivité des alliages, et établit une règle de somme qui relie les déphasages au niveau de Fermi à la charge de l'impureté. Cette règle de somme joua plus tard un rôle essentiel dans le cas d'électrons fortement corrélés, notamment dans l'effet Kondo. Une autre découverte importante fut celle des oscillations de Friedel, responsables par exemple de la formation de structures magnétiques incommensurables. On montre comment elles peuvent être déduites de diverses méthodes : de la théorie des collisions, de la théorie des perturbations, d'approximations self-consistentes ou de la méthode des fonctions de Green. Si la théorie des collisions ne tient pas compte de la structure électronique, et par conséquent de la structure de bandes, ces effets peuvent facilement être inclus dans d'autres théories, par exemple en faisant appel à l'approximation des liaisons fortes.

\* Corresponding author.
*E-mail address:* jvillain@infonie.fr (J. Villain).

## 1. Introduction

The language of physicists evolves. In this issue of the *C. R. Physique* devoted to Jacques Friedel, articles by physicists of various generations and various sub-specialities coexist, and their language is different. This makes communication with non-specialists and students difficult. The aim of the present introduction is to bridge the gap between experts and newcomers, to allow students to read old textbooks, which may be very useful because they were written by those who discovered the phenomena and often understand them better because they remember the difficulties they met, the mistakes which should be avoided, the approximations that failed and those that succeeded. We try to present a synthetic view of a few essential concepts of metal physics.

## 2. Simple mechanisms in pure metals

The physics of metals, where electrons are not localized, but itinerant, is a difficult matter. In the middle of the twentieth century, a few physicists tried to define basic concepts that can bring a better understanding. Among them, Jacques Friedel's part was particularly important.

The task was facilitated by previous discoveries, for instance the *Jahn–Teller distortion*, discovered in 1937 [1], which will now be briefly recalled [2]. Assume a highly symmetric system (a complex or an impurity in a crystal) where an electronic state is (for instance twice) degenerate and occupied by a single electron. The Jahn–Teller theorem states that this high symmetry state is unstable and distorts. The reason is that a weak distortion (or any small perturbation) always splits the degenerate levels into one which is at higher energy and one which is at lower energy (Fig. 1a). The electron goes into the lowest state, and thus the energy is lowered by the distortion. The Jahn–Teller distortion occurs whenever an electronic state has a $n$-fold degeneracy and is occupied by less than $n$ electrons. An example (Fig. 1b) is given by an octahedral complex in which the central ion has a single valence electron on a p-shell. The electron has to choose between 3p orbitals that are directed along the three axes of the octahedron. If it chooses for instance the $z$ axis, the distance $a$ between the ligands on this axis is different from the distance $b$ between the ligands on axes $y$ and $z$.

The Jahn–Teller distortion occurs in small complexes or in insulators containing impurities, but similar effects occur in metals because electronic states just above and just below the Fermi surface are *nearly* degenerate. An example is the *Peierls distortion* [3–5]. It is an instability of a one-dimensional conductor formed by electrons interacting with regularly spaced ions. The ions lower the electronic energy if they modulate their distance with a period $\pi/k_F$, where $k_F$ is the Fermi wave vector of the electron gas. The reason is the following: the distortion mixes electronic states of wave vectors $k$ and $k - 2k_F$. If $k$ is very slightly larger than $k_F$, then $k - 2k_F$ is very slightly larger than $-k_F$ and the corresponding energies $\epsilon(k)$ and $\epsilon(k - 2k_F)$ are almost equal. The distortion lowers the lower energy and increases the higher level, but since there is no electron on this level, the total electronic energy is lowered.

The Peierls distortion (Fig. 2) is a property of one-dimensional conductors (which make them insulating or semi-conducting!). But similar effects occur in certain three-dimensional metals. It has even been suggested [5] that the crystal structure of many covalent materials may be considered as resulting from a Peierls distortion of a fictitious metal. In subsection 2.2, a distortion particular to certain three-dimensional crystal structures will be addressed.

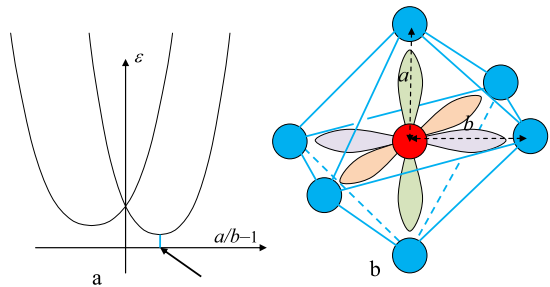

**Fig. 1.** The Jahn–Teller effect. a) Energy of an electron as a function of the distortion $a/b$ of the surrounding. If there are two (or more) degenerate states in the highly symmetric state, a slight distortion increases one of the energies and lowers the other one(s). If there is a single electron, its energy is therefore lowered, b) Example of an ion in an octahedral environment with a partly occupied p shell.

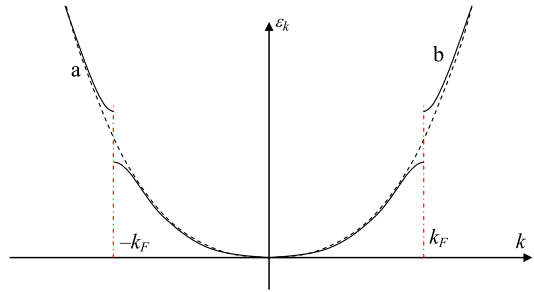

**Fig. 2.** The Peierls distortion of a one-dimensional conductor. A lattice modulation of period $2k_F$ lowers the energy of the electrons below the Fermi level and increases the energy above the Fermi level. The total electronic energy is therefore lowered.

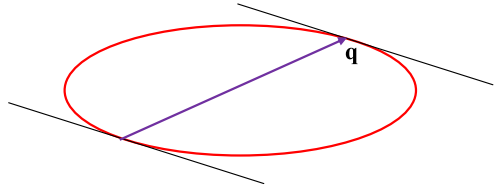

**Fig. 3.** If **q** joins two points of the Fermi surface where the tangent planes are parallel, the phonon frequency $\omega_{ph}(\mathbf{q})$ has a Kohn anomaly for this value of **q**. 'Nesting' occurs when large parts of the Fermi surface are very close to the tangent planes.

### 2.1. Kohn anomaly

The phonon frequency $\omega_{ph}(\mathbf{q})$ of a metal may have anomalies for certain values of the wave vector **q**, which are related to the Fermi surface [6]. Such anomalies are called Kohn anomalies.

To introduce the subject, we choose the case of a very anisotropic conductor, which is nearly one-dimensional, but not sufficiently to undergo the Peierls transition. With appropriate, but not essential simplifying assumptions, the positions $x_i$ of the ions (assumed identical with mass $m$) obey the linearized equations of motion

$$m \, \mathrm{d}^2 x_i / \mathrm{d}t^2 = \sum_j A_{ij}(x_j - x_i) \quad (1)$$

so that the Fourier transform $x_q$ obeys

$$m \, \mathrm{d}^2 x_q / \mathrm{d}t^2 = A(\mathbf{q}) x_q \quad (2)$$

The quantity $A(\mathbf{q}) = \sum_j A_{ij} \exp[i\mathbf{q}\cdot(R_j - R_i)]$ thus appears to be the restoring force of the oscillator. The phonon frequency $\omega_{ph}(\mathbf{q})$ is readily deduced from (2), namely

$$\omega_{ph}^2(\mathbf{q}) = A(\mathbf{q})/m \quad (3)$$

When the Peierls transition is approached (for instance by lowering the temperature), the distorted phase approaches stability and therefore the restoring force $A(\mathbf{q})$ approaches 0 for $q_x = 2k_F$, where $x$ is the direction of high conductivity. Therefore the phonon frequency $\omega_{ph}(\mathbf{q})$ approaches 0 too. As a function of **q**, it has a minimum for $q_x = 2k_F$.[1] Such anomalies of quasi-one-dimensional conductors have been extensively investigated in Orsay [4]. More generally, Walter Kohn [6] proposed to use the phonon spectrum (which can be obtained from inelastic neutron scattering) to obtain "images of the Fermi surface". He noted that, if **q** joins two points of the Fermi surface where the tangent planes are parallel, then the gradient of $\omega_{ph}(\mathbf{q})$ is infinite (Fig. 3). However, such anomalies are generally very difficult to observe. In practice, Kohn anomalies often designate any anomaly of the phonon frequency that is due to conduction electrons. It can be for instance a minimum, or a discontinuity of the slope (Fig. 4). Kohn anomalies can be observed if the Fermi surface has an appropriate shape allowing for the property of *nesting*. This means that, if the Fermi surface is translated by a certain vector **Q**, large portions of the translated surface almost coincide with large portions of the original Fermi surface. Palladium is an example [7].

In a simple theoretical treatment [4], the effect of conduction electrons is considered as a perturbation of the ion–ion interaction. The resulting correction to formula (3) is an additive term proportional to the Lindhard function (see formula (18) below), which gives the response of an electron gas to a perturbation. A difficulty arises from the fact that the Lindhard function depends on the frequency $\omega$ of the perturbation. The appropriate value of $\omega$ would be the phonon frequency $\omega_{ph}(\mathbf{q})$, whose self-consistent determination is not possible analytically. This problem can often be solved by replacing $\omega$ by 0 in (18). This is the *adiabatic approximation*, often justified by the fact that phonon frequencies are generally small with respect to electronic frequencies.

---

[1] Such soft phonons can also appear in certain materials without any effect of conduction electrons, e.g., in ferroelectrics just above the transition.

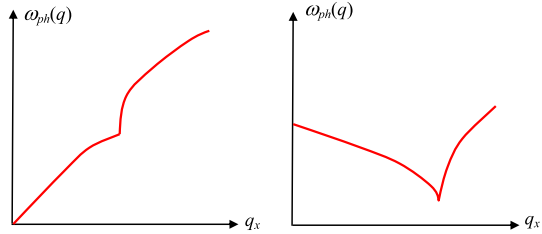

**Fig. 4.** Typical phonon dispersion curves which are usually considered as possible Kohn anomalies in metals.

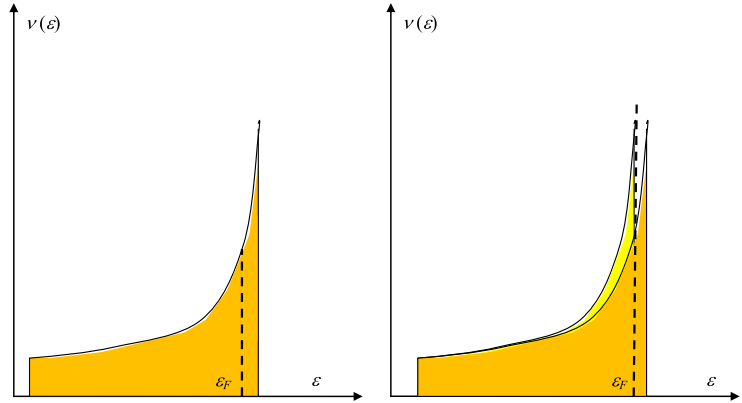

**Fig. 5.** Labbé–Friedel distortion. a) In the cubic structure, there exist three families of conducting chains in orthogonal directions that have the same density of electronic states with a singularity just above the Fermi level. b) In the presence of a tetragonal distortion, the singularities occur at different locations for the three chains, so that the states corresponding to the singularity in the density of states of one or two chains can be completely occupied.

### 2.2. The Labbé–Friedel distortion

A transition from a cubic structure to a tetragonal one is observed at low temperature in intermetallic materials of formula $V_3X$ (X = Si, Ga, Ge, Sn, etc.) or $Nb_3Sn$. Labbé and Friedel [8] explained this distortion as a kind of collective Jahn–Teller effect. V or Nb atoms form conducting chains that have three perpendicular directions. If interchain conduction is neglected, the density of states $\nu(\epsilon)$ diverges at the top of the band and at the bottom. Indeed the energy of an electron of wave vector **k**, counted from the bottom of the conduction band, is $\epsilon(k) = \hbar k^2/m$, where $m$ is the effective mass. Thus $k(\epsilon)$ is proportional to $\sqrt{\epsilon}$ whereas $\nu(\epsilon)$, which for a one-dimensional system is proportional to $dk/d\epsilon$, is proportional to $1/\sqrt{\epsilon}$. In the compounds addressed here (called A15 by metallurgists), the Fermi level is close to the singularity. If a distortion occurs (Fig. 5) the *three* families of chains have their singularities at different locations, and one or two of the chains may have the whole singularity below the Fermi level, which implies a large energy gain. The electronic energy of the other chains or chain is increased, but this energy loss is smaller than the energy gain. The detailed calculation [8] is more complicated than for the true Jahn–Teller effect, but the analogy is clear: the two energy levels of the Jahn–Teller effect are just replaced by two singularities.

The Peierls distortion and the Labbé–Friedel distortion are properties of certain pure metals that arise from a simple mechanism. Similarly, simple concepts have been proposed for dilute alloys, or of impurities in metals. A pioneer of this conceptualization was Mott [9], but considerable advances have been made by Friedel and his collaborators, as recalled by Georges [10].

In the next sections, certain properties of dilute alloys, or of impurities in metals, will be summarized.

### 3. Impurities in metals and dilute metallic alloys

Dilute alloys were for physicists the source of various problems, especially the three following ones.

i) The electrical resistivity does not vanish at low temperature, in contrast with that of pure metals. At temperature $T = 0$, there is a residual resistivity proportional to the impurity concentration. What is this resistivity?

ii) There is often an electric charge localized on the impurity. This charge is screened by conduction electrons, so that the electric field vanishes far from the impurity. How is the screening charge distributed? What is the resulting electric potential?

iii) There may also be a magnetic moment localized on the impurity. This moment is, so to speak, screened by the spins of the conduction electrons, so that the ground state is a singlet, i.e. a non-magnetic state. This is the Kondo effect.

*3.1. Impurities in metals: a scattering phenomenon*

The electrical resistivity of alloys at low temperature results from the scattering of conduction electrons by impurities. Scattering of free particles is a well-known problem of quantum mechanics [11], and Friedel had the idea to apply the standard formalism to alloys.

Let this standard formalism be recalled. The simplest case is when the scattering potential has spherical symmetry. In the absence of scattering, the electronic wave function would be[2]

$$\psi_k^0(\mathbf{r}) = \exp(i\mathbf{k}\cdot\mathbf{r}) \tag{4}$$

If a scattering potential with spherical symmetry around the origin O is introduced, the wave function can be expanded in a series of Legendre polynomials $P_\ell(\cos\theta)$, where $\theta$ is the angle between vectors **k** and **r**. The scattering potential will be assumed to vanish beyond a certain distance $a$. Then it can be shown [11] that the wave function has the following form for $r > a$,

$$\psi_k(\mathbf{r}) = \sum_{\ell=0}^{\infty}(2\ell+1)i^\ell \exp(i\delta_\ell)[j_\ell(kr)\cos\delta_\ell - n_\ell(kr)\sin\delta_\ell]P_\ell(\cos\theta) \tag{5}$$

where $j_\ell(u)$ and $n_\ell(u)$ are respectively spherical Bessel and Neumann functions (actually just polynomials in $1/u$, $\sin u$ and $\cos u$). The quantities $\delta_\ell(k)$ depend on the scattering potential, they are real and are called *phase shifts*. Expression (5) should be valid even when the scattering potential vanishes, and therefore also applies to the plane wave (4). In that case the phase shifts vanish:

$$\psi_k^0(\mathbf{r}) = \exp(i\mathbf{k}\cdot\mathbf{r}) = \sum_{\ell=0}^{\infty}(2\ell+1)i^\ell j_\ell(kr)P_\ell(\cos\theta) \tag{6}$$

Formula (5) can also be written as

$$\psi_k(\mathbf{r}) = \exp(i\mathbf{k}\cdot\mathbf{r}) + (1/r)f_k(\theta)\exp(ikr) \tag{7}$$

with

$$f_k(\theta) = (1/k)\sum_{\ell=0}^{\infty}(2\ell+1)\sin\delta_\ell P_\ell(\cos\theta)\exp(i\delta_\ell) \tag{8}$$

If $ka \ll 1$, the phase shifts of order $\ell > 0$ are negligible, and the scattered wave reduces to its term $\ell = 0$, so that it has spherical symmetry because $P_0(u) = 1$.

Phase shifts can also be defined in two-dimensional conductors. In one-dimensional conductors, there is a single phase shift $\delta_0$.

*3.2. Resistivity of a dilute alloy*

Using formula (8), Friedel and his collaborators were able to calculate the residual resistivity of a dilute alloy, i.e. the resistivity at temperature $T = 0$. They derived the formula [12,13]

$$\Delta\rho = \frac{4\pi e^2 c}{Nm\Omega k_F}\sum_\ell \ell \sin^2(\delta_{\ell-1} - \delta_\ell) \tag{9}$$

where $c$ is the impurity concentration, $\Omega$ is the atomic volume, $e$ the electronic charge, $m$ the electronic mass and $N$ the number of electrons per unit volume. Phase shifts were calculated representing each impurity (i.e. each atom of the minority element) by a potential depending on a parameter $\kappa$: for instance screened Coulomb potential decaying as $\exp(-\kappa r)/r$. Then the phase shifts should be calculated as functions of $\kappa$. Finally $\kappa$ should be determined. This determination can be done using Friedel's sum rule, which will be discussed in the next subsection.

Using formula (9), Faget de Castelnau and Friedel [12] obtained a good agreement with experimental results for Zn, Ga, Ge and As impurities in copper (Fig. 6).

---

[2] Vectors are designated by bold characters, except in indices when there is no confusion. A normalization constant is omitted in formula (4).

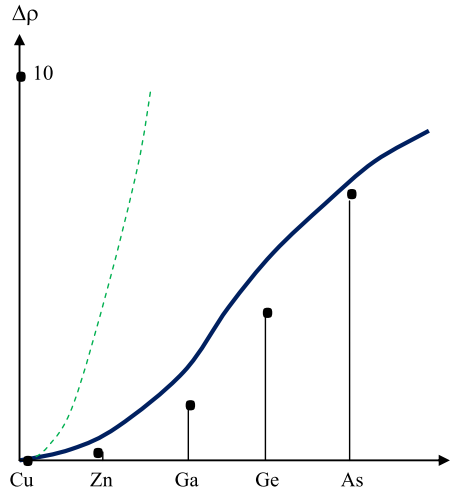

**Fig. 6.** Residual resistivity of various impurities in copper according to the theory of Faget de Castelnau and Friedel [12] (thick, continuous curve) compared with experimental points and a previous theory of Mott and Jones [9] (dashed line). The unit is the microohm·cm per %.

### 3.3. Friedel's sum rule

As an effect of the electron-impurity potential, a certain (average) number $\Delta n$ of conduction electrons are usually attracted to the impurity site. This number may be very close to an integer but, since the medium is a metal with itinerant electrons, it is usually not an integer.

Friedel's sum rule [14] gives $\Delta n$ as a function of the phase shifts, namely

$$\Delta n = \frac{2}{\pi} \sum_\ell (2\ell + 1)\delta_\ell(k_F) \tag{10}$$

The power of this rule lies in the fact that, most of the phase shifts often vanish. An extreme case is when only one phase shift (usually $\delta_0$) is different from 0. Then Friedel's sum rule contains no summation! This is the case in Nozières' treatment of the Kondo effect [15] (see section 9 below and reference [10]).

## 4. Friedel oscillations: the three-dimensional case

### 4.1. Screening

The Coulomb potential is long ranged. However, in a metal or in an electrolyte, two distant charges do not exert an appreciable force upon each other because other charges screen the Coulomb force. The question is: how is the screening charge distributed? If this charge is carried by a classical fluid, the screened electric potential and the screening charge at distance $r$ from a charged impurity are found to decrease exponentially with $r$, as $\exp(-\kappa r)/r$. This result is an element of the Debye–Hückel theory of electrolytes [16]. The argument is as follows: the electric potential $V(r)$ and the screening charge $\rho(r)$ at distance $r$ of the screened charge are related by the Poisson equation $\nabla^2 V(r) = -\rho(r)/\epsilon_0$. On the other hand, the probability of having a charge at point **r** is proportional to $\exp[-\beta V(\mathbf{r})]$ according to Boltzmann's statistics ($\beta = 1/k_B T$, $k_B =$ Boltzmann's constant, $T =$ temperature). At long distances, $V$ is small, so that $\rho(r)$ is just proportional to $-V(r)$ and the Poisson equation reduces to $\nabla^2 \rho(r) = \kappa^2 \rho(r)$, where $\kappa^2$ is proportional to $1/T$. The solution is $\rho(r) \sim \exp(-\kappa r)/r$.

To describe the screening of a charged impurity in metals, the first attempt was to apply an approximation proposed by Thomas [17] and by Fermi [18] in 1927. In the Thomas–Fermi theory, the screened potential $\delta V(r)$ and the screening charge $\delta \rho(r)$ are postulated to be proportional, as they are in electrolytes. As a result, the decay is again exponential, $\rho(r) \sim \exp(-\kappa r)/r$, but, in contrast with electrolytes, $\kappa$ is independent of temperature.[3]

However, the assumption of a local proportionality relation $\delta \rho(r) \sim \delta V(r)$ is acceptable only for localized charges, not for itinerant electrons that propagate nearly freely with an almost constant velocity. This was realized in the middle of the last century, when Friedel and his coworkers showed that the decay is oscillatory. The validity of the Thomas–Fermi approximation was then discussed more systematically by Hohenberg and Kohn [19], who showed that it correctly describes long-wavelength charge variations only.

---

[3] The screening length $1/\kappa$ is obtained by writing that the Fermi level $\epsilon_F$ is the same everywhere. Thus, if there is a local excess potential $\delta V$ that shifts the electronic energy by $e\delta V$, the electrons of energy higher than $\epsilon_F$ should go away. Their number per unit volume is $e\delta V D(\epsilon_F)$ where $D(E)$ is the electronic density of states (number of electronic states per unit volume and energy unit). For free electrons, $D(E) = (2m^3 E)^{1/2}/(2\pi^2\hbar^3)$. Thus $\delta\rho(r) = e^2 D(\epsilon_F)\delta V(r)$ and $\kappa^2 = e^2 D(\epsilon_F)/\epsilon_0$.

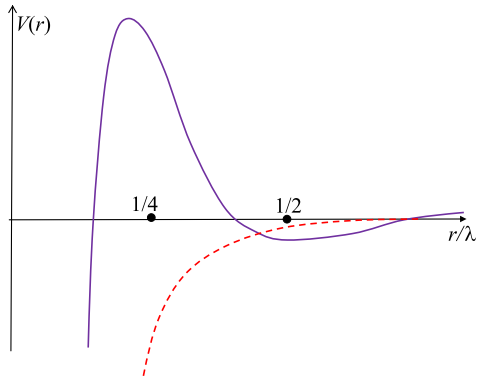

**Fig. 7.** A typical screened Coulomb potential showing Friedel oscillations (full curve) compared with the Thomas–Fermi approximation (dotted curve).

Friedel oscillations can be derived by various methods. We shall first give Friedel's original argument, or rather a simplified version in which all phase shifts are assumed to vanish except $\delta_0$. The general treatment just involves more cumbersome equations.

### 4.2. Friedel oscillations and phase shifts

In Friedel's first papers, whose historical importance is recalled by Daniel [20], the crystal structure was neglected. Then, in the absence of impurity, the electronic states are plane waves characterized by their wave vector **k**. In the presence of a (single) impurity, electronic wave functions $\psi_k(\mathbf{r})$ have the form (5), but they can still be characterized by the wave vector **k** of the incident wave. The electronic density is the sum of the contributions $|\psi_k(\mathbf{r})|^2$ of the various wave functions. Subtracting the electronic density $|\psi_k^0(\mathbf{r})|^2$ in the absence of impurity, one obtains the excess electronic density corresponding to the wave vector **k**, namely

$$\Delta_k(\mathbf{r}) = |\psi_k(\mathbf{r})|^2 - |\psi_k^0(\mathbf{r})|^2 = [j_0(kr)\cos\delta_0 - n_0(kr)\sin\delta_0]^2 - j_0^2(kr)$$

or

$$\Delta_k(\mathbf{r}) = [n_0^2(kr) - j_0^2(kr)]\sin^2\delta_0 - 2j_0(kr)n_0(kr)\cos\delta_0\sin\delta_0$$

Remembering that $j_0(u) = \sin u/u$ and $n_0(u) = -\cos u/u$, one obtains

$$\Delta_k(\mathbf{r}) = (kr)^{-2}[\cos(2kr)\sin\delta_0 + \sin(2kr)\cos\delta_0]\sin\delta_0 \tag{11}$$

It is of interest to define the local density of states $g(\mathbf{r}, \epsilon)$, defined for non-interacting electrons by

$$g(\mathbf{r}, \epsilon) = \sum_k |\psi_k(\mathbf{r})|^2 \delta(\epsilon - \epsilon_k) \tag{12}$$

where $\psi_k(\mathbf{r})$ is the normalized wave function of an electron with wave vector **k** and $\epsilon_k$ is its energy. In the absence of impurity, $g(\mathbf{r}, \epsilon)$ is independent of **r**. An impurity at the origin modifies $g(\mathbf{r}, \epsilon)$ by an amount $\Delta g(\mathbf{r})$. Assuming that $\epsilon_k$ depends only on $|\mathbf{k}|$, relation (11) yields

$$\Delta g(\mathbf{r}) = (A/r^2)\sin[2k(\epsilon)r + \varphi] \tag{13}$$

where $2k(\epsilon)$ is the inverse function of $\epsilon(k)$, $A$ is a constant and $\varphi = \delta_0$. Fig. 7 shows typical Friedel oscillations compared with the Thomas–Fermi approximation.

The local density of states at the Fermi level can be measured by a scanning tunnel microscope [21,22] (STM). The case of the (111) surface of Cu is particularly spectacular because it has an electronic surface state whose Fermi "surface" is nearly a circle, so that the two-dimensional version of (13) (with $1/r$ instead of $1/r^2$) holds. The complicated Fermi surface of bulk copper plays no role. It is worth saying a few words about the remarkable versatility of the STM, which is able to observe the surface atoms, but also (by adjusting the voltage) to forget them and focus on electronic states.

The excess electronic density $\Delta\nu(r)$ can be obtained by integrating (13) over the energy after multiplying by the Fermi function (i.e. the occupation probability of each state). At zero temperature, this means that (13) should be integrated over $k < k_F$. Friedel oscillations are again obtained, but the dependence in $r$ is different, namely $1/r^3$ instead of $1/r^2$ [23]. Indeed, for large $r$, (13) is the derivative with respect to $k_F$ of

$$\nu(r) = -\frac{B}{r^3}\cos(2k_F r + \varphi) \tag{14}$$

where $B$ is a constant. Formulae (13) and (14) are still valid for large $r$ if higher phase shifts are taken into account [24]. The oscillating function (14) is quite different from the exponential behavior predicted by the Thomas–Fermi approximation.

## 5. Phenomena related to Friedel oscillations

In the previous sections, a single impurity was considered. If there are two impurities, each of them perturbs the electron gas, the two perturbations interact, and an interaction between impurities results, which is an oscillating function of their distance. This effect is particularly important for atoms deposited on a surface, since their mobility is larger than in the bulk, and the oscillating interaction influences the position of the impurities. This effect has been observed by scanning tunnel microscopy [25].

If the impurity is a magnetic one, the interaction is between the magnetic moments of the impurities. The magnetic moments can be nuclear spins [26] or electronic ones [27,28]. The resulting Rudermann–Kittel–Kasuya–Yosida interactions (RKKY) are oscillating functions of the distance, alternatively ferro- and antiferromagnetic, and they are responsible for the spin glass structure of certain alloys [29].

Actually the magnetic moments do not need to be those of impurities. Oscillating interactions also exist in pure metals and are responsible for helical or modulated magnetic structures of Cr and of certain rare earths.[4]

## 6. A more realistic approach: the tight-binding approximation

The success of the calculations described in sections 3 and 4 is rather surprising, since the true electronic wave functions in the pure metal are very different from the assumed plane wave $\exp(i\mathbf{k}\cdot\mathbf{r})$. The true wave functions are Bloch waves, $u_k(\mathbf{r})\exp(i\mathbf{k}\cdot\mathbf{r})$, where $u_k(\mathbf{r})$ is a periodic function, but in the neighborhood of each atom at point $\mathbf{R}$, the Bloch function is not very different from a wave function $\varphi(\mathbf{r}-\mathbf{R})$ of the free atom. For instance, in sodium, the wave functions of conduction electrons locally look like 2s orbitals. Thus, the electrons can often be assumed to be tightly bound to the neighboring atom, so that Bloch wave functions of the pure metal have the form

$$\psi_k(\mathbf{r}) = N^{-1/2} \sum_{\mathbf{R}} \varphi(\mathbf{r}-\mathbf{R})\exp(i\mathbf{k}\cdot\mathbf{R})$$

where $N$ is the number of lattice sites. The lattice is assumed to contains a single atom per unit cell. The factor $N^{-1/2}$ ensures normalization. Assuming a single, non-degenerate orbital $\varphi$ per site, ignoring the spin and neglecting interactions between electrons, the Hamiltonian can be written as

$$\mathcal{H} = \sum_{\mathbf{RR}'} t_{\mathbf{RR}'} c_{\mathbf{R}}^+ c_{\mathbf{R}'} \tag{15}$$

where $c_{\mathbf{R}}^+$ and $c_{\mathbf{R}}$ are electron creation and destruction operators on the atom $\mathbf{R}$. The constants $t_{\mathbf{RR}'}$ are called *resonance integrals*. Indeed their evaluation requires the integration of an expression that depends on the wave functions $\varphi(\mathbf{r}-\mathbf{R})$ and $\varphi(\mathbf{r}-\mathbf{R}')$ and on the electron–ion interaction.

A difficulty arises from the fact that the atomic wave functions $\varphi(\mathbf{r}-\mathbf{R})$ and $\varphi(\mathbf{r}-\mathbf{R}')$ at two neighboring sites $\mathbf{R}$, $\mathbf{R}'$ have no reason to be orthogonal. Theorists sometimes ignore this problem and assume orthogonality.

Formula (15) holds in a pure metal. In an alloy, the different energies of electrons on different sites should be taken into account and the tight-binding[5] Hamiltonian reads

$$\mathcal{H} = \sum_{\mathbf{RR}'} t_{\mathbf{RR}'} c_{\mathbf{R}}^+ c_{\mathbf{R}'} + \sum_{\mathbf{R}} \epsilon_{\mathbf{R}} c_{\mathbf{R}}^+ c_{\mathbf{R}} \tag{16}$$

Moreover, interactions between electrons may be taken into account. The popular Hubbard model introduces a repulsive (Coulomb) interaction between two electrons on the same site only. Since there is a single orbital per atom, the two electrons necessarily have opposite spins. The Hubbard model is the simplest model of itinerant magnetism. If one wants to analyze the effect of a single magnetic impurity (e.g., the Kondo effect), it is sufficient to take into account the Coulomb interaction on the impurity site. This is the Anderson model, addressed by Georges [10] and in section 9 of this paper.

Thus, the electronic states of interest are generated by the action of operators $c_{\mathbf{R}}$ and $c_{\mathbf{R}}^+$ on the ground state $|0\rangle$. These operators destroy or create a conduction electron at site $\mathbf{R}$. Their Fourier transforms $c_{\mathbf{k}}$ and $c_{\mathbf{k}}^+$ are also useful. The direct space is no longer considered continuous as in the previous sections, but discrete.

In many cases, the conduction band consists of several orbitals that should be labeled by an additional index $\alpha$. For instance, in a transition metal as Fe, the conduction bands are the 4s and 3d bands. And of course, each orbital can be occupied by two electrons with opposite spins.

Applications of the tight-binding approximation to bulk metals and their surfaces are reviewed in references [5,10,30,31].

---

[4] Rare earths and chromium are somewhat different. In rare earth, magnetism is due to localized, 4f electrons, and the oscillating interaction results from the action of conduction electrons. In chromium, the magnetism itself results from conduction electrons.

[5] This terminology means that ion–electron interactions are strong, not the chemical binding.

## 7. Perturbative and self-consistent treatments

As seen above, the original derivation of Friedel oscillations ignored the consequences of the crystal structure and the properties of Bloch waves. A possible alternative method is just perturbation theory. If a weak potential $V_q(\omega)\exp[i(\mathbf{q}\cdot\mathbf{r} - \omega t)]$ is applied to the gas of Bloch electrons (whose interactions are neglected), the perturbation of the electronic density is $v_q(\omega)\exp[i(\mathbf{q}\cdot\mathbf{r} - \omega t)]$ where

$$v_q(\omega) = e\chi(\mathbf{q},\omega)V_q(\omega) \tag{17}$$

where the so-called Lindhard function has been introduced,

$$\chi(\mathbf{q},\omega) = \frac{1}{N\Omega}\sum_k \frac{f_0(\epsilon_{\mathbf{k+q}}) - f_0(\epsilon_{\mathbf{k}})}{\epsilon_{\mathbf{k+q}} - \epsilon_{\mathbf{k}} - \hbar\omega - i\eta} \tag{18}$$

where $\Omega$ is the volume of the unit cell and $f_0(\epsilon) = 1/[1 + \exp(\epsilon - \epsilon_F)/k_BT]$ is the Fermi function at temperature $T$. The quantity $\eta$ is real, positive and very small. It appears because the perturbation is assumed to be switched on from a time $t = -\infty$ at a very slow rate $\eta$, corresponding to an imaginary frequency. It avoids any divergence of (18) for real frequencies.

In the case of interest here, $\omega = 0$ and the impurity potential has many Fourier components $V_q$, the effect of which is additive.

The band structure, and therefore the crystal structure, are taken into account through the function $\epsilon_k$. The explicit calculation can be done numerically for each particular material. We only note that, in a three-dimensional metal, formula (14), and therefore Friedel oscillations, can be derived [32] at temperature $T = 0$ from (17) and (18) if $\epsilon_k$ is assumed to depend only on $|\mathbf{k}|$. If $\epsilon_k$ has a weak angular dependence (as it really has), the behavior of the electron density at long distances from the impurity may be modified, but Friedel oscillations persist, as will be argued in subsection 8.4. Of course, if the energy $\epsilon_k$ and the potential $V_k$ are known, the response (17) can be calculated numerically.

In first order perturbation theory, formulae (17) and (18) are valid for any metal, including one- and two-dimensional ones. For a one-dimensional metal, $\chi(q,0)$ diverges as $\ln(1/|q - 2k_F|)$ near $2k_F$. This corresponds to Friedel oscillations of the form $(1/r)\cos(2k_F r + \varphi)$ in the charge density.

### 7.1. Self-consistent treatment

Actually equations (17) and (18) are not limited to the linear response approximation and may be regarded as a self-consistent approximate solution, provided $V_q$ is not the applied potential $V_q^0$, but takes screening into account. In three dimensions, this implies [33] that

$$V_q(\omega) = \left[1 - \frac{4\pi e^2}{q^2}\chi(\mathbf{q},\omega)\right]^{-1}V_q^0(\omega) \tag{19}$$

## 8. Green's functions

In the current literature, it is usual to express physical quantities in terms of Green's functions. They are derived from correlation functions such as $\langle c_R^+(t)c_{R'}\rangle = \langle\exp(it\mathcal{H})c_R^+\exp(-it\mathcal{H})c_{R'}\rangle$, where $\mathcal{H}$ is the Hamiltonian. We shall see that Green's functions provide a more general way to obtain Friedel oscillations. The temperature will be assumed to be 0.

In cases of interest here, electrons do not interact. One can introduce the one-electron states and the corresponding creation and destruction operators $c_\alpha^+$ and $c_\alpha$. Then the Hamiltonian can be written as

$$\mathcal{H} = \sum_\alpha \hbar\omega_\alpha c_\alpha^+ c_\alpha \tag{20}$$

The local operators $c_R$ can be written as linear combinations of the eigenmode operators $c_\alpha$:

$$c_R = \sum_\alpha \psi_\alpha(\mathbf{R})c_\alpha \tag{21}$$

The correlation functions $\langle c_R^+(t)c_{R'}\rangle$ are therefore linear combinations of the functions

$$\langle 0|c_\alpha^+(t)c_\alpha|0\rangle = \langle 0|c_\alpha^+ c_\alpha|0\rangle \exp(i\omega_\alpha t) \tag{22}$$

where the temperature has been taken equal to 0. It is convenient to multiply this expression by the Heaviside function[6] $\theta(t)$, so that the Fourier transform reads

---

[6] $\theta(t) = 0$ for $t < 0$, $\theta(t) = 1$ for $t \geq 0$.

$$\int_{-\infty}^{\infty} \langle 0 | \theta(t) c_\alpha^+(t) c_\alpha | 0 \rangle \exp(-i\omega t) dt = \int_0^{\infty} \langle 0 | c_\alpha^+(t) c_\alpha | 0 \rangle \exp(-i\omega t) dt = \frac{1}{i(\omega_\alpha - \omega)} \langle 0 | c_\alpha^+ c_\alpha | 0 \rangle \quad (23)$$

This expression vanishes if $\hbar\omega_\alpha$ is higher than the Fermi energy $\hbar\omega_F$. Indeed, $c_\alpha | 0 \rangle = 0$ in that case. Since we would not like to lose the information above $\hbar\omega_F$, it is the right time to introduce the Green's functions

$$\hat{G}_{RR'}(t) = i\theta(t) \langle c_R^+(t) c_{R'} \rangle - i\theta(-t) \langle c_{R'} c_R^+(t) \rangle \quad (24)$$

which are linear combinations of

$$\hat{G}_\alpha(t) = i\theta(t) \langle c_\alpha^+(t) c_\alpha \rangle - i\theta(-t) \langle c_\alpha c_\alpha^+(t) \rangle \quad (25)$$

whose Fourier transform $G_\alpha(\omega)$ can be obtained at temperature $T = 0$ by addition of (23) and an analogous term:

$$G_\alpha(\omega) = \int_{-\infty}^{\infty} \hat{G}_\alpha(t) \exp(-i\omega t) \, dt = \frac{1}{\omega - \omega_\alpha} \quad (26)$$

It follows from (26) and (21) that

$$G_{RR'}(\omega) = \sum_\alpha \psi_\alpha^*(R) \psi_\alpha(R') \frac{1}{\omega - \omega_\alpha} \quad (27)$$

which can be written in a concise form as

$$G_{RR'}(\omega) = \langle 0 | c_R^+ \frac{1}{\omega - \mathcal{H}/\hbar} c_{R'} | 0 \rangle + \langle 0 | c_{R'} \frac{1}{\omega - \mathcal{H}/\hbar} c_R^+ | 0 \rangle \quad (28)$$

### 8.1. Green's functions and local density of states

The local density of states (12) can be obtained from (27) if the frequency has a complex value $\omega - i\eta$. We define

$$G_{RR'}^-(\omega) = G_{RR'}(\omega - i\eta) = \sum_\alpha \psi_\alpha^*(R) \psi_\alpha(R') \frac{1}{\omega - \omega_\alpha - i\eta} = \sum_\alpha \psi_\alpha^*(R) \psi_\alpha(R') \frac{\omega - \omega_\alpha + i\eta}{(\omega - \omega_\alpha)^2 + \eta^2} \quad (29)$$

A function $G_{RR'}^+(\omega)$ can be defined in the same way, replacing $-i\eta$ by $+i\eta$
In the limit $\eta \to 0$ (29) implies

$$\mathrm{Im} G_{RR'}^-(\omega) = \frac{2}{\pi} \sum_\alpha \psi_\alpha^*(R) \psi_\alpha(R') \delta(\omega - \omega_\alpha) \quad (30)$$

This formula is of particular interest for $R = R'$. Then, it defines the local density of states $g_R(\omega)$, already introduced by (12), which is a special form of (30) when the eigenfunctions of the Hamiltonian can be characterized by a wave vector **k**. Thus

$$g_R(\omega) = \frac{2}{\pi} \mathrm{Im} G_{\mathbf{R},\mathbf{R}}^-(\omega) \quad (31)$$

### 8.2. Green's functions in a pure crystal

Relation (31) shows that Green's functions have a physical significance. But how can they be calculated? How can they be used to derive Friedel oscillations around an impurity? The first thing to do is to calculate them in the pure metal. Then the eigenstates $\alpha$ are Bloch functions characterized by their wave vector **k**. The Green's function $G_{RR'}^0(\omega) = G^0(\mathbf{r}, \omega)$ only depends on $\mathbf{r} = R - R'$, and (29) reads (omitting the tilde)

$$G_0^-(\mathbf{R} - \mathbf{R}', \omega) = (1/N) \sum_k \frac{1}{\omega - \omega_k - i\eta} \exp[i\mathbf{k}(\mathbf{R} - \mathbf{R}')] \quad (32)$$

where $N$ is the number of atoms (or unit cells, because we assume one atom per unit cell). The factor $1/N$ arises from the normalization condition of the wave functions, and ensures that the Green's function has a finite value for an infinite sample, since the summation is over $N$ vectors. Assuming spherical symmetry of the function $\omega_k$, one obtains in three dimensions

$$G_0^-(\mathbf{r}, \omega) = \frac{2\pi a^3}{r} \int \frac{\sin kr}{\omega - \omega_k - i\eta} k \, dk$$

where $a^3$ is the volume of the unit cell. At long distance $r$, the sine oscillates so rapidly that the only important contribution comes from $k$ very close to $k(\omega)$, the inverse function of $\omega(k)$. It follows that

$$G_0^-(\mathbf{r}, \omega) \simeq \frac{2\pi^2 k(\omega)}{\alpha r}[\cos k(\omega)r + i \sin k(\omega)r] \tag{33}$$

where $\alpha = d\omega/dk$.

*8.3. Green's functions and scattering*

A link between Green's functions and the scattering problem may be obtained as follows. For an incident wave $\Psi_k^0(\mathbf{R}) = \exp(i\mathbf{k} \cdot \mathbf{R})$ and a scattering potential $V_R$, the outgoing wave is [34]:

$$\Psi_k(\mathbf{R}) = \Psi_k^0(\mathbf{R}) - \frac{1}{4\pi} \sum_{R'} G_0^+(\mathbf{R} - \mathbf{R}', \omega) V_{R'} \Psi_k(\mathbf{R}') \tag{34}$$

This formula holds in the tight binding approximation with one orbital per site. A more general formula has been given for instance by Blandin [34].

Another property of Green's functions is that they determine the wave function at all lattice sites **R** when particles of energy $\hbar\omega$ are constantly created at $\mathbf{R} = 0$. Indeed the Schrödinger equation writes $\sum_{R'} \mathcal{H}_{RR'} \Psi(R') - \hbar\omega\Psi(R) = \alpha\delta_{0R}\Psi(0)$, where $\alpha$ is a constant. Inversion of this relation yields, using (28),[7]

$$\Psi(\mathbf{R}) = G_0(\mathbf{R}, \omega)\Psi(0) \tag{35}$$

Equation (34) can be solved by iteration. At first order, $\Psi_k(\mathbf{R}')$ can be replaced by $\Psi_k^0(\mathbf{R}')$. From the iteration series emerges the T-matrix $T_{RR'}(\omega)$ defined by the relation

$$G(\mathbf{R}, \mathbf{R}', \omega) = G^0(\mathbf{R} - \mathbf{R}', \omega) + \sum_{R''} G^0(\mathbf{R} - \mathbf{R}'', \omega) T_{\mathbf{R}'',\mathbf{R}'''}(\omega) G^0(\mathbf{R}''' - \mathbf{R}', \omega) \tag{36}$$

This formula yields the Green's function in the presence of an impurity. The $T$-matrix is usually short ranged, so that, at long distance, (13) can be derived from (36), (33) and (31).

*8.4. Long-distance behavior of the oscillations in a real crystal (with a non-spherical Fermi surface)*

Although the Green functions, the charge density, and the local density of states can be in principle calculated from (36), explicit formulae as (13) have only been derived assuming that the electron energy $\hbar\omega_k$ depends only on $|\mathbf{k}|$, which is a completely unphysical assumption. The Green's functions of a real, perfect crystal have been discussed by Blandin [34], on the basis of an apparently unpublished thesis of Laura Roth [35]. However, a detailed calculation was published by Roth et al. a few years later [36].

In the simplest cases, the long-distance behavior is described in a pure crystal by a formula analogous to (33), i.e. the Green's function decays as $\cos(Kr - \varphi)/r$ in any given direction, but $K$ depends on the direction of $\mathbf{r}$, and of course also on $\omega$. More precisely, $K$ is the value of $k$ at the points where the normal to the surface $\omega(\mathbf{k}) = \omega$ is parallel to $\mathbf{r}$. Of course there are at least two such points, $\mathbf{k}$ and $-\mathbf{k}$. But there may be more points if the surface $\omega(\mathbf{k}) = \omega$ is not convex. In that case the Green's function decays as a linear combination of functions of the form $\cos(Kr - \varphi)/r$, with different values of $K$ for a given direction $\mathbf{r}/r$. Fig. 8 shows a case with two vectors $K_1$ and $K_2$.

In practice, the long-distance behavior is not extremely important, since the main physical effects result from the first oscillations. However, it is often convenient to write analytic formulae as (13) and (14), and it is good to evaluate their validity. For instance, (13) is often correct, except that $k(\epsilon)$ should be replaced by a different value $k(\epsilon, \mathbf{r}/r)$ in each direction.

Knowing the Green's function of the pure crystal, one can deduce the local density of states around an impurity from (31) and (36). An integration then gives the charge density. Details can be found in articles by Gautier [37] and by Roth et al. [36]. Since several propagation vectors $K_1$, $K_2$... may be present, depending on the direction $\mathbf{R}/R$, Friedel oscillations may be much less regular than implied by (14), and observed by STM in Cu(111) and in graphene [38].

## 9. Kondo effect: phase shifts and quasiparticle Fermi liquid theory

Following the ideas and theories developed by Jacques Friedel and collaborators to describe local impurities in metallic alloys with the introduction of the concept of virtual bound states and phase shifts experienced by the wavefunction of the conduction electron when scattered off the impurity, P.W. Anderson introduced the following model Hamiltonian [39], $H_{\text{And}}$

$$H_{\text{And}} = H_{\text{d}} + H_{\text{c}} + V \tag{37}$$

---

[7] Relation (35) may help to understand the relation with Green's functions used in the theory of differential equations, or in acoustics and elasticity.

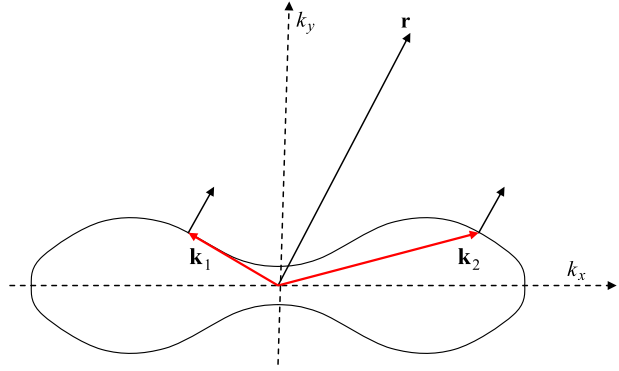

**Fig. 8.** If the surface $\omega(k) = \omega$ is not convex, it may be normal to $\mathbf{r}$ in two points $\mathbf{k}_1$ and $\mathbf{k}_2$ (or more). In the case shown by the figure, the Green's function $G_0(\mathbf{r}, \omega)$ decays at long distance $r$ as $[A_1 \cos(\mathbf{k}_1 \cdot \mathbf{r} - \phi_1) + A_2 \cos(\mathbf{k}_2 \cdot \mathbf{r} - \phi_2)]/r$, where $A_1, \phi_1, A_2, \phi_2$ depend on $\omega$ and $\mathbf{r}/r$.

with

$$H_d = \sum_\sigma \varepsilon_0 d_\sigma^\dagger d_\sigma + U n_\uparrow n_\downarrow$$

$$H_c = \sum_{k,\sigma} \varepsilon_{k\sigma} c_{k\sigma}^\dagger c_{k\sigma}$$

$$V = \sum_{k,\sigma} (t_\sigma c_{k\sigma}^\dagger d_\sigma + h.c.)$$

where $H_d$ describes the localized electrons in the impurity site with spin $S = 1/2$ interacting among themselves through the repulsive Coulomb interaction, $U$; $H_c$ describes the conduction electrons with spin $s = 1/2$ in the metallic host, and $V$ the hybridization between the localized and conduction electrons. $n_\sigma = d_\sigma^\dagger d_\sigma$ is the number of localized electrons for spin $\sigma$, $\varepsilon_0$ their energy (in the presence of a magnetic field $B$, $\varepsilon_0$ should be replaced by $\varepsilon_\sigma = \varepsilon_0 + \sigma g \mu_B B/2$ resulting from Zeeman splitting), and $t_\sigma$ the hybridization matrix element between $|k\sigma\rangle$ and $|\sigma\rangle$ states, assumed to be $k$-independent. At equilibrium, the energies of the electrons in the metal are distributed according to the Fermi–Dirac distribution function: $n_F(\varepsilon_{k\sigma}) = [\exp[(\varepsilon_{k\sigma} - \mu)/k_B T)] + 1]^{-1}$, where $\mu$ is the chemical potential in the metal ($\mu$ is taken as the origin of energy in the following).

In its original form or considering its further extensions, this model provides the basis to describe correlated metals or superconductors with the fundamental issue about the existence of a Mott metal–insulator transition induced by correlations [40], as well as heavy-fermion systems [41] in which instead of having one isolated impurity, one has a periodic array of "impurities" (anomalous rare-earth or actinide atoms), with the possible onset of quantum phase transitions[8] and non-Fermi liquid behavior observed in the vicinity of a quantum critical point. This model has experienced a resurgence of interest these last two decades, since it is believed to be the appropriate model to describe the quantum dot [42–45], in which the central region of the dot is assimilated to the impurity, and the two leads – left and right – constitute the two parts of the metallic reservoir. Interestingly, the quantum dot can be driven in non-equilibrium conditions when for instance a dc bias voltage $V$ applied between the two leads sets a difference between the chemical potentials of each lead.

In the absence of Coulomb interaction ($U = 0$), the Hamiltonian is solvable and leads to the formation of Friedel's virtual bound state with the density of states of localized electrons of spin $\sigma$, $A_{d\sigma}(\omega)$ given by

$$A_{d\sigma}(\omega) = \frac{1}{\pi} \frac{\Gamma_\sigma}{(\omega - \varepsilon_0)^2 + \Gamma_\sigma^2} \tag{38}$$

where $\Gamma_\sigma = \pi \sum_k |t_\sigma|^2 \delta(\varepsilon_0 - \epsilon_{k\sigma})$ (equal to $\pi |t_\sigma|^2 \rho_{0\sigma}$ in the limit when the conduction electron band is of infinite width, leading to a constant density of states for spin $\sigma$ given by $\rho_{0\sigma}$). The expression (38) corresponds to a Lorentzian peak centered at the energy $\varepsilon_0$ of the impurity level, of width $\Gamma_\sigma$ resulting from the virtual excursion or transfer of the localized electron into the conduction metallic host given by Fermi's golden rule. This peak is precisely the virtual bound state introduced by Jacques Friedel. It is easy to check that the phase shift at $\omega = 0$, $\delta(0)$ experienced by the wavefunction of the conduction electron when scattered off the impurity, is directly related to the number of electrons of spin $\sigma$ in the impurity site, $n_{d\sigma} = \int_{-\infty}^0 A_{d\sigma}(\omega) d\omega$ according to

$$\delta_\sigma(0) = \pi n_{d\sigma} \tag{39}$$

---

[8] A quantum phase transition is a phase transition which can occur even at zero temperature, the order being destroyed by quantum fluctuations.

corresponding to Friedel's sum rule as outlined in Section 3.3. This result reflects the complete screening of the positive charge of the impurity by the negative charge of the conduction electrons surrounding the impurity site.

The opposite limit ($V = 0$) corresponds to the atomic limit. The total number of electrons, $n_\text{d}$ in the impurity site varies with the energy of the impurity level, $\varepsilon_0$ from $n_\text{d} = 2$ (doubly-occupied site) when $\varepsilon_0 = -U$, to $n_\text{d} = 0$ (empty site) when $\varepsilon_0 = 0$. An important case is the particle-hole symmetric limit corresponding to $\varepsilon_0 = -U/2$; In this limit, the total density of states of localized electrons has two peaks, located at $\omega = \varepsilon_0$ and $\omega = \varepsilon_0 + U$, symmetric to each other with respect to $\omega = 0$, leading to $n_\text{d} = 1$ (singly-occupied site).

The question then is to understand the effect of the Coulomb interactions, $U$ on this virtual bound state picture. Remarkably, in the presence of interactions, the phase shift at the Fermi level, $\delta(0)$, is related to the number of electrons of spin $\sigma$ in the impurity site, $n_{\text{d}\sigma}$, by exactly the same above-mentioned Friedel's sum rule in the absence of Coulomb interaction – see Eq. (39). The proof was given by Langreth [46] making use of general Ward identities. Friedel's sum rule is completely general whatever the occupancy of the impurity level and the strength of the interactions are.

An important step further was taken by Schrieffer and Wolff [47], who showed that in the so-called local moment limit $|\varepsilon_0| \gg \Gamma$ and $|\varepsilon_0 + U| \gg \Gamma$ ($n_d \simeq 1$), the Anderson model maps at small energy into the Kondo model defined by the Hamiltonian, $H_K$ (within a scattering potential term, which is omitted)

$$H_K = H_\text{d} + \frac{1}{2} J \vec{S}(0) \cdot \sum_{k,k',\sigma,\sigma'} c^\dagger_{k\sigma} \vec{\sigma}_{\sigma\sigma'} c_{k'\sigma'} \tag{40}$$

where $J = |t_\sigma|^2 \left\{ \frac{1}{\varepsilon_d} - \frac{1}{\varepsilon_d + U} \right\}$, $\vec{S}(0) = \frac{1}{2} \sum_{\sigma,\sigma'} d^\dagger_\sigma \vec{\sigma}_{\sigma\sigma'} d_{\sigma'}$ is the spin of the impurity, and $\vec{\sigma}$ are the Pauli matrices. $J$ is positive in the local moment regime considered and corresponds to antiferromagnetic coupling between the two spins. This model gives rise to the Kondo effect at low temperature, corresponding to the dynamical screening of the impurity spin by the conduction electron spin. J. Kondo [48] developed a perturbation theory in $J$ for this model. He obtained a logarithmic increase in the resistivity with decreasing $T$, explaining the experimental observations. However the calculations also lead to unphysical divergence of the resistivity at low temperature, meaning that the perturbation theory in $J$ ceases to be valid when $T$ goes lower below the Kondo temperature,[9] the characteristic low-energy scale of the Kondo model. Below $T_K$, the perturbation theory in $J$ breaks down and the system crosses over to another regime where one has to use alternative non-perturbative methods. What happens below $T_K$ was first foreseen by P.W. Anderson by devising a perturbative renormalization group method that he named as Poor Man's Scaling [50], which consists in perturbatively eliminating excitations to the edges of the noninteracting band. This method indicated that, as temperature is decreased, the effective coupling between the spins of the impurity and of the conduction band electron, $J_\text{eff}$, increases to $+\infty$, leading to the formation of a singlet state between the spin of the impurity and the spin of the conduction electrons in the impurity site. This argument was pursued in a more rigorous way by K.G. Wilson [49] using the Numerical Renormalization Group (NRG) method. Wilson was able to show that when the energy scale goes to zero, the distribution of eigenstates becomes similar to what would prevail if $J$ were going to $+\infty$. Therefore, even if the bare $J$ is small, there is a smooth crossover from the weak coupling to the strong coupling regime, $J = +\infty$ as $T$ goes below $T_K$. The real breakthrough made by NRG calculations is to describe the whole crossover from the weak coupling regime to the strong coupling regime when $T$ or $B$ decreases.

Soon after Wilson's paper, it was realized [15] by Philippe Nozières that the "low temperature-end of the story" can be described as a local Fermi liquid. Let us remind that the concept of Fermi liquid theory introduced by Landau in 1957–1959 is based on the assumption that the system exhibits an adiabatic continuity when the interactions are turned on adiabatically, and then there is a one-to-one correspondence between the single-particle excitations of the gas of noninteracting particles (i.e. the bare electron) and the single-particle excitations of the gas of interacting particles at sufficiently short time scales (defining the "quasiparticle"). The gas of interacting particles is then described as a system of weakly-interacting "quasiparticles". Nozières formulated a local Fermi-liquid theory for the quasiparticles and expressed it in terms of the phase shift acquired by the quasiparticle when scattered off the Kondo singlet, enlarging Friedel's phase shift concept to the case of correlated impurity. Usually the phase shift depends only on the kinetic energy $\varepsilon$ of the quasiparticle, but here in the spirit of the Landau theory of Fermi liquid, Nozières considered the additional dependence of the phase shift on the distribution function $\delta n_{\sigma'}(\varepsilon')$ of the other quasiparticles with which the quasiparticle interacts:

$$\delta_\sigma(\varepsilon) = \delta_0(\varepsilon) + \sum_{\varepsilon'\sigma'} \varphi_{\sigma,\sigma'}(\varepsilon, \varepsilon') \delta n_{\sigma'}(\varepsilon') \tag{41}$$

Since in the strong coupling regime (low $T$ and $B$), only the states in the vicinity of the Fermi surface matter, one can expand the quantities $\delta_0(\varepsilon)$ and $\varphi_{\sigma\sigma'}(\varepsilon, \varepsilon')$ around the Fermi energy

$$\delta_0(\varepsilon) = \delta_0 + \alpha\varepsilon + \beta\varepsilon^2 + \ldots \tag{42}$$

$$\varphi_{\sigma,\sigma'}(\varepsilon, \varepsilon') = \varphi_{\sigma,\sigma'} + \psi_{\sigma,\sigma'}(\varepsilon + \varepsilon') + \ldots \tag{43}$$

---

[9] In Kondo's work, $T_K$ is identified as the temperature at which, in the perturbation theory in $J$, the third-order term of the transmission equals the second-order term. $\rho_0$ is the total density of states of conduction electrons at the Fermi level. A more precise result is $T_K = (JD)^{1/2} \exp -1/(J\rho_0)$ [49].

At low $T$ and $B$, one can keep only the first three and two terms in the r.h.s. of the previous two expressions, respectively. Therefore, only four quantities enter the expression of the phase shift: $\delta_0$, $\alpha$, $\varphi^s$ and $\varphi^a$ (where $\varphi^s$ and $\varphi^a$ are defined from: $\varphi_{\sigma,\pm\sigma} = \varphi^s \pm \varphi^a$). In the local moment regime considered, the symmetric part $\varphi^s$ does not play any role and only the asymmetric part $\varphi^a$ matters. Setting $m = \delta n_\uparrow - \delta n_\downarrow$, where $\delta n_\uparrow = \sum_{\varepsilon'} \delta n_{\sigma'}(\varepsilon')$, one gets

$$\delta_\sigma(\varepsilon) = \delta_0 + \alpha\varepsilon + \sigma\varphi^a m \qquad (44)$$

According to Friedel's sum rule, $\delta_0 = \pi/2$ in the local moment regime considered when $n_\sigma = 1/2$. $\alpha$ and $\varphi$ can be viewed as phenomenological Fermi-liquid parameters. They correspond respectively to energy-dependent elastic scattering and local interaction between the quasiparticles resulting from the polarization of the Kondo singlet complex. They can be related to the zero-temperature impurity spin susceptibility, $\chi_s$, the inverse of which defines the Kondo temperature, $T_K$, the characteristic low-energy scale of the Kondo model. Their exact values can be extracted from the results obtained by Wilson using Numerical Renormalized Group method. Indeed, by resorting simple physical arguments as the fact that the Kondo singularity is tied to the Fermi energy, it can be shown that $\alpha$ and $\varphi$ are connected through the relation

$$\alpha - 2\rho_0\varphi^a = 0 \qquad (45)$$

Using this quasiparticle Fermi liquid theory, Nozières was able to derive the low-$T$ and -$B$ physical properties of the model. He recovers the anomalous Wilson ratio $R$ (the dimensionless ratio of the spin susceptibility to the linear coefficient of the specific heat) with a value $R = 2$, and predicts a quadratic dependence of the resistivity in $T$ and $B$, characteristic of a Fermi-liquid behavior, with the determination of the coefficients of the $T^2$ and $B^2$ terms.[10]

Independently, Yosida and Yamada [52] developed a diagrammatic Fermi-liquid theory based on perturbation calculations in successive orders in $U$. These calculations show that the ground state of the Anderson model is a Fermi liquid in all parameter regimes whatever the occupancy of the impurity level and the strength of the interactions are. Their calculations, first performed at the second order in $U$, reproduce the main results obtained by the quasiparticle Fermi liquid theory. By using Ward identities, they were able to show that the results hold even up to infinite order in $U$. This type of theory is at the origin of renormalized perturbation theory [53,41] as reviewed in Georges' contribution [10].

Both of these quasiparticle and diagrammatic Fermi-liquid theories proved to be extremely useful. They could be applied to various extensions of the Kondo or Anderson model. They have the great advantage to be applicable to out-of-equilibrium situations, as for instance when a finite bias voltage is applied between the left and right leads of a quantum dot as mentioned at the beginning of this section. This is quite remarkable given that the alternative methods to treat the Kondo model or Anderson model at intermediate or strong $U$ – based on either Numerical Group Renormalization, or Bethe Ansatz, or conformal field theory – are not easily transposable to the situation of non-equilibrium.

This section highlights the importance of Friedel's ideas for the study of correlated impurities as described by the Kondo model or the Anderson impurity model. It turns out that Friedel's sum rule about the phase shift plays a key role in the description of these related scattering problems, triggering the development of Fermi-liquid theories to describe the low-temperature regime of these models. This field has experienced a resurgence of interest during these last two decades with the revival of the Kondo effect in the context of mesoscopic physics.

## 10. Conclusion

The example of Friedel oscillations shows how a phenomenon can be derived from various methods.

Oscillations appear in three different properties:

i) the electronic density (or charge density) around an impurity, given in dimension 3 by formula (14). In dimension 2, the factor $1/r^3$ is replaced by $1/r^2$ and in dimension 1, it is replaced by $1/r$. Graphene is an exceptional case [21,22];

ii) the local density of charge around an impurity, given in dimension 3 by formula (13). In dimension 2, the factor $1/r^2$ is replaced by $1/r$. This quantity is particularly important at the Fermi energy, since it is related to measurable quantities such as the tunnel micrograph;

iii) the one-electron Green function of the pure crystal, given in dimension 3 by (33).

The methods used by Jacques Friedel were not so precise as modern techniques. He was at the origin of many fruitful ideas, but rarely tried to express them in an elaborated form. He did encourage his collaborators to do so, and much has been done in this direction by André Blandin. A great merit of Jacques Friedel is to have been more concerned by the future of his many collaborators than by his own.

---

[10] In a very interesting recent paper, Mora et al. [51] succeeds in extending the quasiparticle Fermi liquid theory to any regime of the Anderson model with the introduction of 4 (instead of 2) Fermi-liquid parameters expressed as a function of the spin and charge susceptibilities (and derivatives) which they again determine from the Bethe Ansatz and Wilson's Numerical Renormalization Group methods.